\documentclass[nalefd, letter]{achemso}
\usepackage{amsmath}
\usepackage{amssymb}
\usepackage{nicefrac}
\usepackage{graphicx}
\usepackage{float}
\usepackage{comment}
\usepackage{color}
\usepackage{soul}
\title{Non-equilibrium surface growth in a hybrid inorganic-organic system}
\author{Nicola Kleppmann and Sabine H. L. Klapp}
\affiliation{
Institut f\"ur Theoretische Physik,
Technische Universit\"at Berlin,
Hardenbergstr. 36,
10623 Berlin,
Germany
}
\email{nicola@kleppmann.de}
\begin{document}

\date{\today}
\begin{abstract}
Using kinetic Monte Carlo simulations, we show that molecular morphologies found
in non-equilibrium growth can be strongly different from those at equilibrium.
We study the prototypical hybrid inorganic-organic system 6P on ZnO(10-10) during thin
film adsorption, and find a wealth of phenomena including re-entrant growth, a critical
adsorption rate and observables that are non-monotonous with the adsorption rate. We identify
the transition from lying to standing molecules with a critical cluster size and discuss
the competition of time scales during growth in terms of a rate equation approach. Our results form a basis 
for understanding and predicting collective orientational ordering during growth in hybrid
material systems.
\begin{figure*}
    \includegraphics{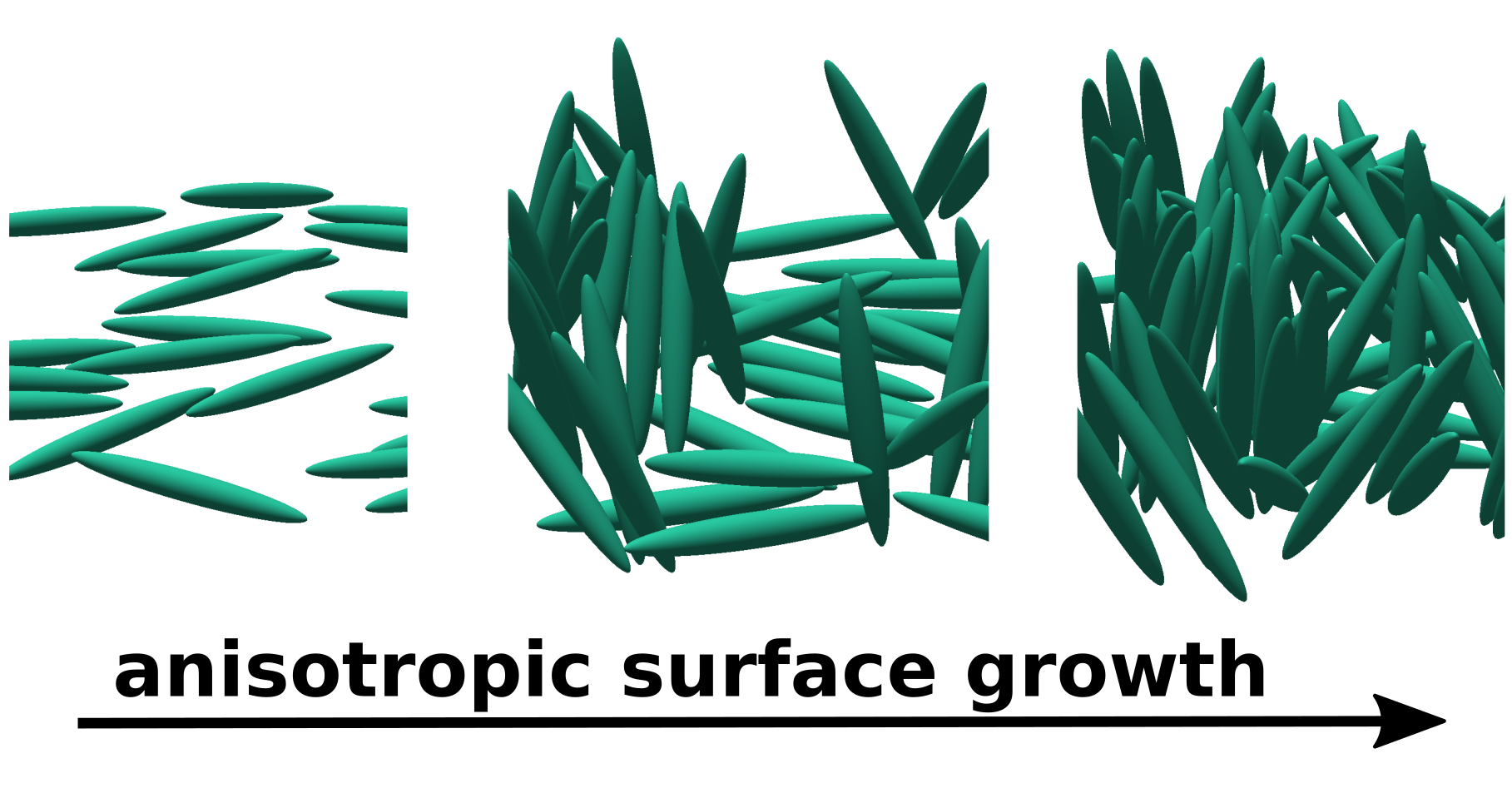}    
\label{fig:thumbnail}
\end{figure*}
\end{abstract}

\maketitle

\begin{figure}
    \includegraphics{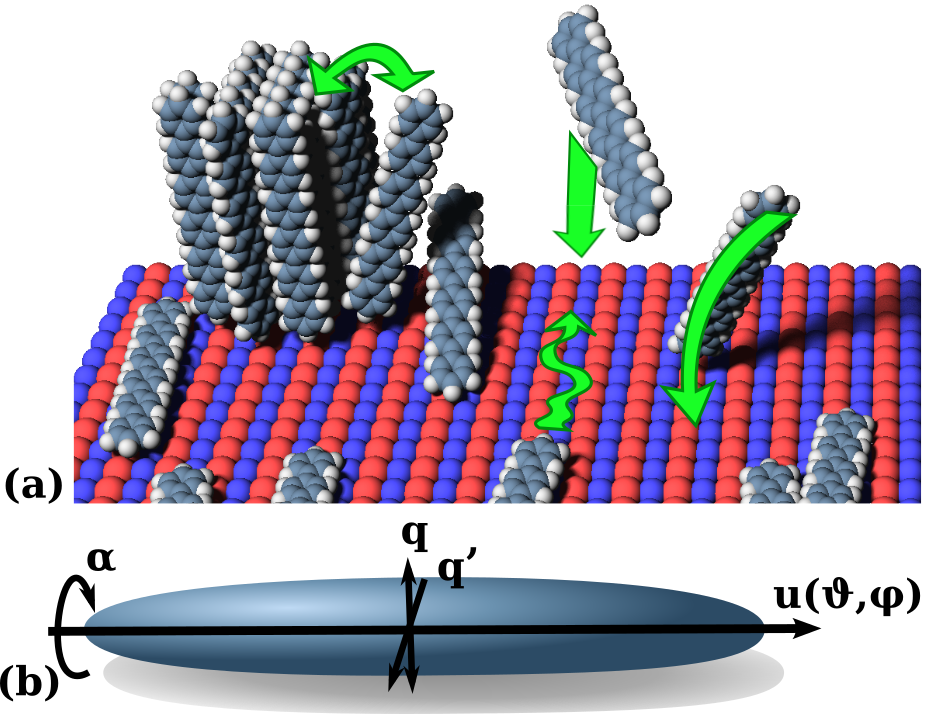}    
\caption{(Color online) Growth schematic for submonolayer growth of 6P on ZnO(10-10). Part (a) depicts the substrate (alternating blue and red colors represent charge lines)
with molecules adsorbing, diffusing translationally and rotationally (see green arrows). Part (b) defines the three principles axes of a molecule, {\bf u}, {\bf q} and {\bf q$\prime$}, as a function of the three Euler angles $\alpha$, $\vartheta$ and $\varphi$ (with {\bf q} and {\bf q$\prime$} defining linear quadrupole moments).
}
\label{fig:1_6P_growth_schematic}
\end{figure}

Hybrid inorganic-organic systems (HIOS) have revolutionized opto-electronic semiconductor technologies 
by combining the high charge carrier densities and high tunability of conjugated organic molecules with
stable, well controlled inorganic substrates \cite{Law2005, Liu2009}. 
At the same time, the design of efficient devices poses fundamental physical questions on a multitude of length- and time-scales, from resonance energy transfer in complex 
environments \cite{Rindermann2011, Blumstengel2006, Vaynzof2012, Xu2013} and energy level alignment \cite{Schlesinger2015, Braun2007}
to classical statistical physics problems such as collective ordering of anisotropic molecules at interfaces, both in equilibrium and during 
non-equilibrium growth, i.e., at finite flux \cite{Hlawacek2013, Zojer2000, Simbrunner2013}. 
 The final orientational ordering is indeed crucial fo HIOS functionality \cite{Hlawacek2013}. Experimentally, 
a number of interesting structural phenomena in HIOS have been observed,
such as the increase of lying clusters with temperature \cite{Blumstengel2010}, the coexistence of domains with 
different molecular tilt angles, and layer-dependent tilt angles \cite{Sparenberg2014}. However, so far there is no
consistent understanding from the theoretical side \cite{Simbrunner2011, Hlawacek2013}. 
A generic, yet unsettled phenomenon seems to be that single molecules lie 
on substrates, while thin films generally form a standing configuration. Understanding this transition is further challenged 
by the fact that in typical HIOS, the substrate generates electrostatic or topological
surface fields with significant impact on molecular ordering \cite{Simbrunner2013}.

In the present letter, we consider a {\em prototypical} HIOS system and ask the question: How does the molecular component transition from lying molecules to 
standing clusters during non-equilibrium, sub-monolayer growth? We demonstrate, using theoretical calculations, that this transition is dominated by an interplay of anisotropic interactions and growth kinetics.

Specifically, we consider sexiphenyl (6P) molecules on a zinc-oxide (ZnO) 10-10 surface, see Fig.~\ref{fig:1_6P_growth_schematic}. This system combines two generic HIOS features: strongly anisotropic molecules with non-symmetric (here dominantly quadrupolar) charge distributions and an electrostatically patterned substrate, here " charge stripes" \cite{Blumstengel2006}. We have recently developed a coarse-grained model \cite{Kleppmann2015} based on DFT parametrization \cite{Sala2011, Golubkov2006},
which reproduces key equilibrium properties.
Here, we employ kinetic Monte Carlo (KMC) simulations to access the large time- and length-scales of collective ordering for different adsorption rates.
Currently, KMC simulations of anisotropic molecules are in their infancy. One conceptual problem is that the molecules'
mobility is influenced by its orientation \cite{Hopp2012, Raut1998, Raut1999}.
In earlier studies of molecular growth, the molecules' orientations are strongly restricted to discrete orientations \cite{Ruiz2004, Choudhary2006, Jana2013} 
or a 2D (stripe-patterned) plane \cite{Hopp2012}. 
However, in reality most organic molecules explore the full, 3D space of orientations \cite{Blumstengel2010, Hlawacek2013}. Here, we consider molecules that are 
translationally confined to a 2D lattice, but have continuous, 3D rotational degrees of freedom.
As very few growth simulations of hybrid systems exist, and since 6P/ZnO(10-10) forms a prototypical case, our results provide first steps towards 
a general understanding of growth phenomena in HIOS.

In KMC simulations, the surface evolution is described as a series of transitions between discrete states, where 
each transition is characterized through a rate. Specifically, we use a `Composition and Rejection' algorithm to select events and determine the system time from event propensities,
a method previously used for biochemical models \cite{Cao2004, Plimpton2009}.
This algorithm is advantageous for simulations with continuous rotational degrees of freedom, as it does not require the computationally expensive 
calculation of a complete process rate catalogue after each process \cite{Chatterjee2007}. 
The processes occuring during surface growth can be summarized in terms of three different types: rotational diffusion (r), translational diffusion (d)
and adsorption (a), described through the rates $r_i^{\textrm{r}}$, $r_i^{\textrm{d}}$ and $r_i^{\textrm{a}}$, respectively (see Fig.~\ref{fig:1_6P_growth_schematic}(a)). 
Each process type is associated with an attempt frequency $\nu$, which we use as propensity in our algorithm. 

The rotational rate allows continuous rotational degrees of freedom. Using an adiabatic approximation, it can be expressed as
\begin{equation}
 r^{\textrm{r}}_i=\nu^{\textrm{r}} \cdot \min \left\{ \exp\left[ \beta\left(H^{\textrm{i}}(i)-H^{\textrm{f}}(i)\right)\right],1\right\}\textrm{,}\label{eq:rot_rate}
\end{equation}
where $H^{\textrm{i}}(i)$ and $H^{\textrm{f}}(i)$ are the effective coarse-grained interaction Hamiltonians for the initial and final configuration of molecule $i$, respectively, 
and $\beta=(k_{\textrm{B}} T)^{-1}$ is defined through the Boltzmann constant $k_{\textrm{B}}$ and the temperature $T$. Here, we set $T=300$\,K. Following \cite{Kleppmann2015}, 
the effective interaction Hamiltonian is written as
\begin{equation}
\label{eq:Heff}
 H_{\textrm{eff}}^{\textrm{pot}}=H_{\textrm{m-m}}+H_{\textrm{m-s}}\textrm{,}
\end{equation}
where $H_{\textrm{m-m}}$ characterizes the intermolecular (6P-6P) interactions
consisting of an electrostatic quadrupolar contribution $V_{\textrm{QQ}}$ \cite{Gray1984} between the
(linear) quadrupoles assigned by \textbf{q} in Fig.~\ref{fig:1_6P_growth_schematic}(b) and a non-electrostatic Gay-Berne interaction $V_{\textrm{GB}}$ \cite{Gay1981, Cleaver1996, Golubkov2006}. 
Correspondingly, the molecule-substrate Hamiltonian $H_{\textrm{m-s}}$ contains a quadrupole-field interaction for a quadrupole denoted by 
\textbf{q}$\prime$ in Fig.~\ref{fig:1_6P_growth_schematic}(b). This electrostatic interaction 
reflects the hybrid nature of the 6P/ZnO(10-10) system.
A single 6P molecule tends to {\em align with} the charge lines of the substrate \cite{Blumstengel2010,Sala2011}.
Further, $H_{\textrm{m-s}}$ includes a non-electrostatic interaction 
$V_{\textrm{LJ}}$ \cite{Hansen1990} that involves attractive $z^{-3}$ interactions between molecule and substrate. The different contributions to $H_{\textrm{eff}}^{\textrm{pot}}$ are parametrized based on 
DFT calculations \cite{Golubkov2006, Sala2011, Kleppmann2015}. 

The rotational rate in Eq.\,(\ref{eq:rot_rate}) lacks transitional information between the initial- and final configuration. This reflects the underlying "adiabatic" 
approximation, i.e., a much faster relaxation of rotational versus translational motion. 
The latter is modelled as an activated process between identical, {\em energetically most favourable} lattice sites of the ZnO(10-10) substrate \cite{Kleppmann2015}, as discussed in detail in 
the SM1. As translational diffusion does not 
include any rotation, the m-s interaction of initial site and destination are identical. Thus, $r_{i}^{\textrm{d}}$ involves
 a constant contribution $E_{\textrm{d}}$ determined through the substrate, as well as the initial neighbour interaction energy $H_{\textrm{m-m}}^{\textrm{i}}$,
\begin{align}
 r^{\textrm{d}}_i=\nu^{\textrm{d}}
 \begin{cases}
  &\min\left(\exp\left[\beta\left(H_{\textrm{m-m}}^{\textrm{i}}(i)-E_{\textrm{d}}\right) \right], 1 \right)\\
  &\qquad \qquad \qquad \qquad \qquad \quad \textrm{if }H_{\textrm{m-m}}^{\textrm{f}} (i) \leq E_{\textrm{d}}\\
  &\min\left(\exp\left[\beta\left(H_{\textrm{m-m}}^{\textrm{i}}(i)-H_{\textrm{m-m}}^{\textrm{f}}(i)\right)\right], 1 \right) \textrm{else}. 
 \end{cases}
\label{eq:diff_rate}
\end{align}
The second case comes into play if the molecule's destination is blocked through repulsive interaction with other molecules, i.e. if $H_{\textrm{m-m}}^{\textrm{f}}(i)>E_{\textrm{d}}$, as discussed in 
SI1.

In principle, information about $E_{\textrm{d}}$ can be obtained from atomistically resolved molecular dynamics (MD) simulations \cite{Palczynski2014}. These have shown that $E_{\textrm{d}}$ is direction-dependent (as one might expect). 
Nevertheless, we here set $E_{\textrm{d}} = 0$\,eV for all directions of diffusion. 
Test calculations for $E_{\textrm{d}} \leq 0.2$\,eV (the free energy barrier found for diffusion along charge lines \cite{Palczynski2014})
indicate that this approximation has no substantial influence on our results.

The final process is the adsorption of molecules, expressed through the adsorption rate
\begin{equation}
 r^{\textrm{a}}_i=\nu^{\textrm{a}}=f\cdot \nu^{\textrm{d}}\textrm{.}\label{eq:ads_rate}
\end{equation}
\begin{figure}
    \includegraphics{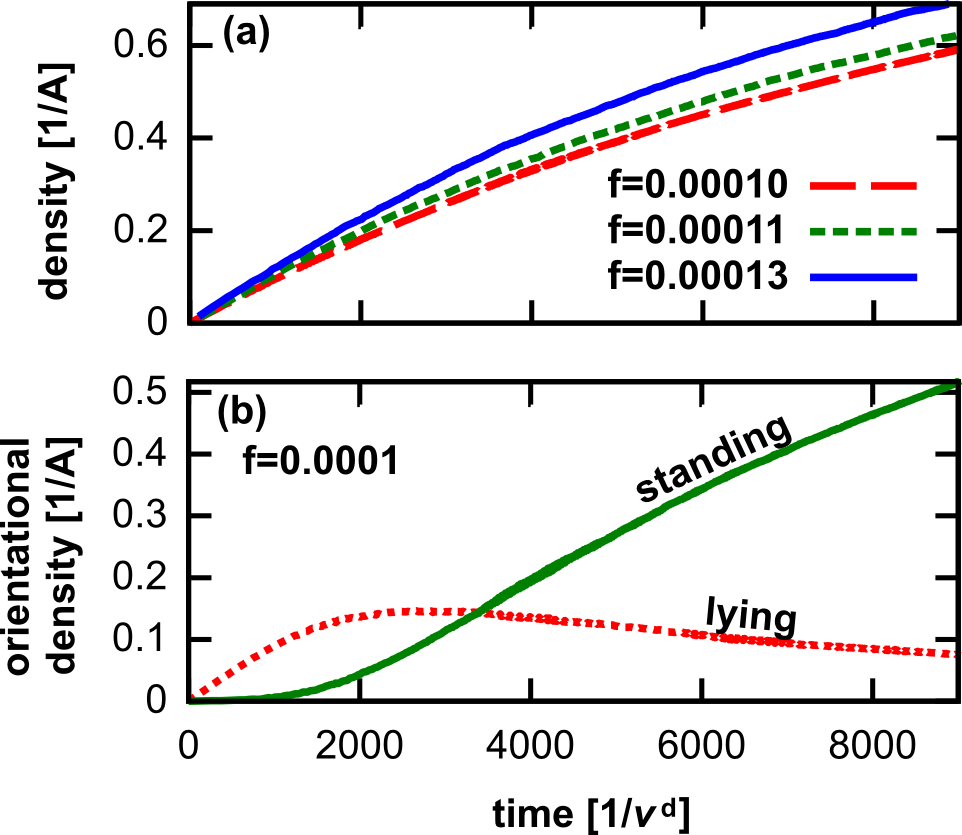}
\caption{(Color online) Growth properties as a function of time for different adsorption rates $f$ . Part (a) depicts the overall density of molecules
while part (b) shows the densities of standing and lying molecules as a function of time. Each molecule with $\left|\vartheta_i \right| \leq 0.25\pi$ is classified as standing.}
\label{fig:2_coverage_part_density}
\end{figure}
\begin{figure*}
    \includegraphics{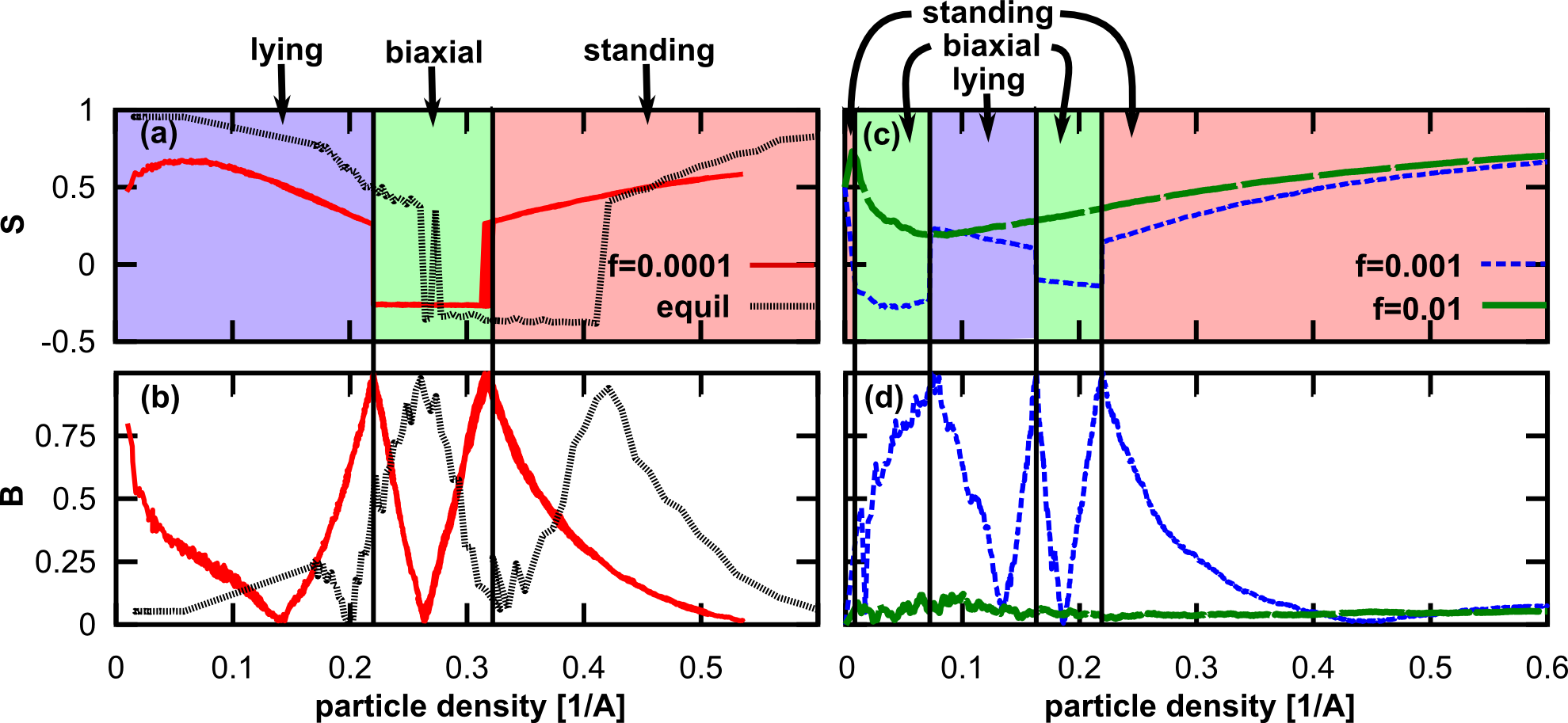}
\caption{(Color online) Orientational order parameters as function of density. (a), (c) nematic order parameter $S$; (b), (d) biaxiality parameter $B$ 
as functions of the molecule density for different $f$. Parts (a) and (c) have underlying colourblocks that mark the different morphologies 
 arising for $f=0.0001$ and $f=0.001$, respectively. The vertical lines mark the transitions between these morphologies. 
 The transitions in equilibrium and at $f=0.01$ are not explicitely marked}.
\label{fig:3_S_B_part_density_all}
\end{figure*}
 Note that we do not consider desorption and only count particles that adsorb on previously {\em unoccupied} substrate sites. The numerical results presented below have been obtained using a vertical adsorption orientation of molecules; the influence of different
adsorption configurations is discussed in SM2. 
We use adsorption rates in the range of $10^{-6}\leq f \leq 1$, which we compare to the upper limit of the free diffusion rate $r^{\textrm{d}}_{\textrm{max}}\approx 2\cdot 10^{9}$\,Hz determined on the basis 
of MD simulations \cite{Palczynski2014}. We find that for $f=10^{-6}$ maximally half a monolayer of molecules adsorbs within $\approx 0.4$\,ms; thus we are far beyond typical time-scales of MD simulations. Simulation details are 
discussed in SM1.

We characterize the orientational order using the order parameter tensor \cite{Allen1989, Strehober2013},
$A_{\alpha\beta}(t)=N^{-1}\sum_{i=1}^N \left\langle u_{i,\, \alpha} u_{i,\, \beta} 
 - \frac{1}{3} \delta_{\alpha\beta}\textrm{Tr}(\textbf{u}_i\otimes \textbf{u}_j) \right\rangle$,
where Tr stands for the trace, $\otimes$ denotes a dyadic product and $\langle\dots\rangle$ is an average over all molecules and all runs in
the system at time $t$.
The nematic order parameter $S \in \left[-0.5,1\right]$ and the biaxiality parameter $B\in \left[0,1\right]$ are derived from the eigenvalues of $A_{\alpha\beta}(t)$ (see SM3).
For a perfectly uniaxal system $S = 1$, $B=0$, while $S = B=0$ represents a completely disordered system. 
Positive values of both, $S$ and $B$, indicate nematic ordering with some degree of biaxiality. Finally, molecular ordering along two orthogonal axes within a plane \cite{Andrienko2002, Andrienko2006} yields
$S=-1/2$, $B=0$, corresponding to maximal biaxiality (for further details, see SM3).

As a starting point, we consider the density of molecules, i.e. the number of molecules per surface unit cell of size $A$, as a function of time in Fig.\,\ref{fig:2_coverage_part_density}\,(a).
The density increases monotonously for all adsorption rates. 
This monotonic, yet non-linear behaviour is consistent with a rate-equation approach introduced below.
In the subsequent plots, we therefore replace the time axis by a density axis (adjusted to the adsorption rate considered).

Figure~\ref{fig:2_coverage_part_density}\,(b) gives first insight into the orientational ordering at small adsorption rates by depicting separately the density of standing and lying molecules.
Initially, the majority of molecules lies on the substrate (with preference of the $x$-direction as supported by the ZnO charge lines), until $t\approx3000\,(\nu^{\textrm{d}})^{-1}$.
Then, the fraction of standing molecules starts to dominate, and at large times (high density) all molecules stand. 
The implications for cluster growth are discussed in SM4, where 
we also estimate the critical cluster size for the lying-standing transition ($\approx 15$ molecules). Both, the lying and the standing configuration correspond to 
equilibrium states (determined at $f=0$ and $r^{\textrm{d}}=0$ by scaling the lattice constants) at low and high densities, respectively \cite{Kleppmann2015}. 
Moreover, the observation that molecules initially lie and later stand during thin film growth closely resembles experimental findings for self-assembled monolayers of decanthiol on silver
\cite{Schreiber1998}: orientationally sensitive scattering measurements show that the first adsorbed molecules nearly immediately form a lying phase,  
which, as the density increases, transforms to a standing phase \cite{Schreiber1998, Schreiber2000, Brutting2006}. This orientational reordering appears to be generic for a wide class of material 
combinations \cite{Schreiber2000, Hlawacek2013, Potocar2011, Brutting2006, Choudhary2006} and 
only very strongly attractive substrates do not support a transition to a standing molecular orientation \cite{Brutting2006}.

\begin{figure}
    \includegraphics{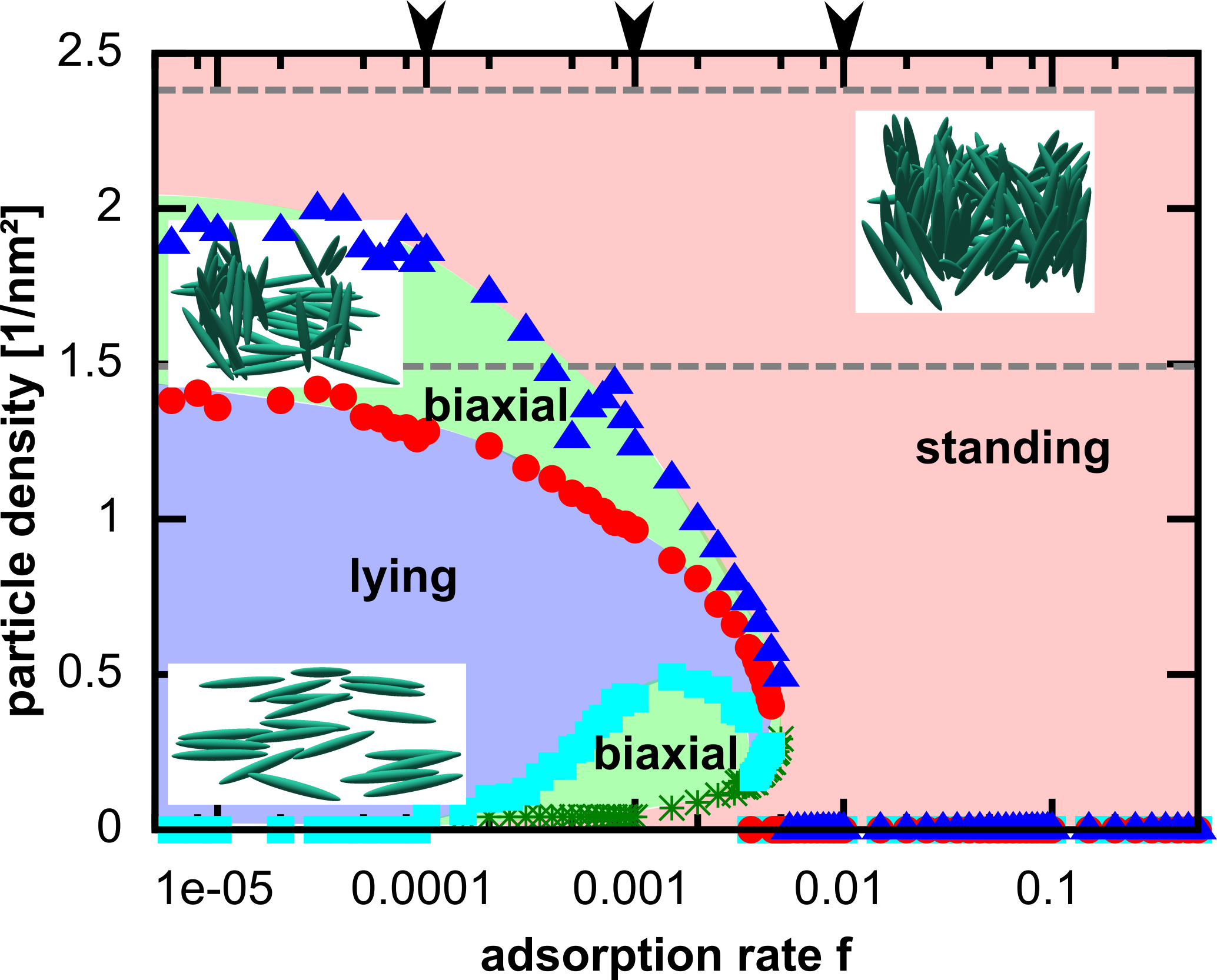}
\caption{(Color online) `Growth state diagram' as a function of adsorption rate and density for
vertically adsorbed molecules. Regions of different orientation are separated through transition points that correspond to a change in 
sign of $S$, as exemplified in Fig.\,\ref{fig:3_S_B_part_density_all}. The dashed gray horizontal lines mark
the transitions expected in equilibrium (see Fig.\,\ref{fig:3_S_B_part_density_all}(b)). 
 Note that the density is here expressed in units of nm$^{-2}$ (contrary to Fig.~\ref{fig:3_S_B_part_density_all}).
The snapshots correspond to the lying, biaxial and standing
configuration for $f=0.0001$, which correspond to molecule densities $0.39$\,nm$^{-2}$, $1.59$\,nm$^{-2}$ and $3.2$\,nm$^{-2}$, respectively. The arrows mark
the adsorption rates shown in Fig.\,\ref{fig:3_S_B_part_density_all}.}
\label{fig:4_transitions_part_A_and_B}
\end{figure}

The dependency of the orientational ordering on the rate of adsorption is analyzed in Fig.\,\ref{fig:3_S_B_part_density_all}, where we plot 
$S$ and $B$ as functions of density for several values of $f$ (including $f=0$). An overview over different $f$ and densities is given in Fig.\,\ref{fig:4_transitions_part_A_and_B}. Apart from statistical fluctuations, the data do not depend on the system size.

In equilibrium, the system displays two transitions (see Fig.\,\ref{fig:3_S_B_part_density_all}(a)): First, from nematic (lying) to a biaxial
phase where the molecules orient either along the charge lines or the $z$-axis (at a density of\,$\approx 0.26\,A^{-1}$) or along the $z$-axis (standing). Second, from
biaxial to full nematic standing at\,$\approx 0.41\,A^{-1}$.

We now turn to non-equilibrium effects. For low, yet non-zero adsorption rates ($f=0.0001$), molecules have less time to form a nematical lying order. 
As a consequence, both the transition from lying to biaxial and the transition from biaxial to 
standing nematic move towards lower densities. 
For very high adsorption rates ($f=0.01$), the standing order initiated through molecule adsorption dominates the 
orientational ordering throughout the growth process, as depicted in Fig.\,\ref{fig:3_S_B_part_density_all}(c). Here, the nematic order parameter $S$
 never assumes negative values and the biaxiality parameter never is significantly larger than zero. 
The intermediate range is dominated by competition of various morphologies, exemplified for $f=0.001$ in Fig.\,\ref{fig:3_S_B_part_density_all} (c).
Initially, the standing orientation dominates. Then, the system transitions to biaxial
at $\approx 0.02\,A^{-1}$ and lying nematic at $\approx 0.078\,A^{-1}$, as the initially adsorbed molecules lie down. With increasing molecule density, 
the molecular morphology observed in equilibrium becomes more significant. Correspondingly, the transition from lying to biaxial, and from biaxial to standing 
resemble the transitions seen for $f=0.0001$ in Fig.\,\ref{fig:3_S_B_part_density_all}(a) and (b), except that they are shifted to lower densities. We call this 
phenomenon `re-entrant' growth, because the initial standing orientation vanishes, but reoccurs during the final stages of growth.\\
The rate-dependency of the collective orientation behaviour is summarized in Fig.\,\ref{fig:4_transitions_part_A_and_B} for a range of adsorption rates and associated densities. 
Here,
three interesting features become apparent. First, there is re-entrant growth for intermediate adsorption rates close to $f=0.001$, as discussed above. Second, the transitions of $S(f)$ (marked 
 in Fig.\,\ref{fig:4_transitions_part_A_and_B}) are non-monotonous with $f$. This observation is verified by multiple runs for $0.005\leq f \leq 0.0005$ for 
lattices with between $2500$ and $10^6$ lattice sites. As a consequence, the orientational ordering at a fixed density of 0.25\,nm$^{-2}$ follows a sequence of states upon increase of $f$: lying, 
biaxial, lying, biaxial and finally standing order. Third, there appears to be a `critical' adsorption rate $f_{\textrm{crit}}\approx0.005$ beyond which no transitions occur. 

To rationalize the occurence of $f_{\textrm{crit}}$, we propose the following rate-equation model involving discretized rotational configurations (for details, see SM5). 
Specifically, we assume that the number of vacant sites, $N_v$, sites occupied by standing particles, $N_s$, and sites occupied by lying particles, $N_l$, change
in time according to
$\dot{N}_v\equiv dN_v/dt=-N_v f$, $\dot{N}_s=N_v f- N_s r$, 
$\dot{N}_l=N_s r$,
where $r$ is the rate of reorientations. The transition from standing to lying is identified with the condition $N_s(t)=N_l(t)$. Solving the rate equations accordingly we find the implicit equation
$(2-\Theta)\,x-2\,x\,\exp[\ln(1-\Theta)/x]=\Theta$
where $\Theta(t)=(N_s(t)+N_l(t))/N$ and $x=f/r$. This yields (see SM5) the critical adsorption rate $f_{\textrm{crit}}^,\approx 0.0055$,
in very good agreemnet with the corresponding simulation result. Physically, our model demonstrates that $f_{\textrm{crit}}$ results from
the competition between the time scales of adsorption ($f^{-1}$) and reorientation ($r^{-1}$).

In summary, our KMC simulations of the prototypical organic-inorganic system 6P on ZnO(10-10) clearly show that non-equilibrium morphologies can be strongly different from those found for $f=0$. The present calculations are based on spatially dependent pair potentials parametrized according to DFT results, and unlike earlier KMC studies \cite{Hopp2012,Raut1998} 
we assume {\em continuous 3D} rotations. This allows us the explore the full complexity of non-equilibrium orientational ordering.

To which extent are the results {\em generic} for hybrid inorganic-organic systems? In fact, while our coarse-grained Hamiltonian (\ref{eq:Heff}) is designed
for a specific material system (and surface temperature $T$), the observed competition between lying and standing orientations
as function of $f$ is a rather universal feature \cite{Schreiber2000}. Regarding surface properties, one would clearly expect an impact
of $T$ and also of the surface structure: For higher $T$ the "orientational bias" exerted by charged stripes of ZnO(10-10) weakens, destabilizing the lying nematic phase already at $f=0$\cite{Kleppmann2015}. In non-equilibrium ($f>0$), we thus expect the critical rate $f_{\textrm{crit}}$ for vertical adsorption (and generally, the regime of rates related to lying phases)
to decrease. For substrate temperatures $T\lesssim 300~K$ \cite{Winkler2016}, the temperature difference between adsorbing molecules and the substrate may cause hot precursor states, which reduce the influence of $T$ \cite{Winkler2013, Morales2014}. Similarly, inorganic surfaces without a biasing field will lead to a destabilization of biaxial phases. Regarding the molecules, decreasing their length (e.g., from 6P to 2P)
reduces the anisotropy of sterical interactions even in the uncharged case \cite{Schreiber2000}; thus, lying phases can exist
up to higher densities \cite{Miriam2016}. For quadrupolar molecules, shorter lengths weaken the electrostatic substrate interactions \cite{Blumstengel2006}, so we again expect $f_{\textrm{crit}}$ to decrease.

The understanding of these different, nonequilibrium molecular structures forms an important ingredient for later calculations of optical behaviour. 
Hence, our understanding of the collective ordering during growth of hybrid structures contributes to the creation of greener, more sustainable and cost-efficient opto-electronic 
devices \cite{Kagan1999, Raimondo2012, Roberts2008}. We hope that our simulation results stimulate further experiments in this direction.\\
This work was supported by the Deutsche Forschungsgemeinschaft within CRC 951 (project A7). 
We acknowledge interesting discussions with M.~Klopotek, F.~Schreiber, and M.~Oettel.
%

\end{document}